\documentclass[conference]{IEEEtran}
\usepackage{graphicx}
\usepackage[sort&compress, numbers]{natbib}
\usepackage{verbatim}
\usepackage[]{algorithm2e}
\usepackage{graphicx}
\usepackage{booktabs}
\usepackage{relsize}
\usepackage{colortbl}
\usepackage{enumerate}
\usepackage{amsmath}
\usepackage{tikz}
\usepackage{pgf}
\usepackage{pbox}
\usepackage{rotating}
\usepackage{multirow}
\usepackage{balance}
\usepackage{subfigure}
\usepackage{xcolor}
\usepackage{authblk}
\usepackage{cleveref}[2012/02/15]
\usepackage{arydshln}
\crefformat{footnote}{#2\footnotemark[#1]#3}

\usepackage{soul}

\renewcommand\refname{REFERENCES}

\renewcommand{\labelitemi}{--}
\newcommand{\startlist}{\begin{list}{\labelitemi}{\leftmargin=1em}\setlength{\itemsep}{-1mm}}
\newcommand{\stoplist}{\end{list}}

\begin{document}
\title{Bug or Not?\\ Bug Report Classification using N-Gram IDF}

\author[1]{Pannavat Terdchanakul}
\author[1]{Hideaki Hata}
\author[2]{Passakorn Phannachitta}
\author[1]{Kenichi Matsumoto}
\affil[1]{Graduate School of Information Science, Nara Institute of Science and Technology, Nara, Japan}
\affil[2]{College of Arts, Media and Technology, Chiang Mai University, Chiang Mai, Thailand}
\affil[1]{\{pannavat.terchanakul.pp5, hata, matumoto\}@is.naist.jp}
\affil[2]{passakorn.p@cmu.ac.th}

\maketitle

\begin{abstract}
Previous studies have found that a significant number of bug reports are misclassified between bugs and non-bugs, and that manually classifying bug reports is a time-consuming task. To address this problem, we propose a bug reports classification model with N-gram IDF, a theoretical extension of Inverse Document Frequency (IDF) for handling words and phrases of any length. N-gram IDF enables us to extract key terms of any length from texts, these key terms can be used as the features to classify bug reports. We build classification models with logistic regression and random forest using features from N-gram IDF and topic modeling, which is widely used in various software engineering tasks. With a publicly available dataset, our results show that our N-gram IDF-based models have a superior performance than the topic-based models on all of the evaluated cases. Our models show promising results and have a potential to be extended to other software engineering tasks.
\par Keywords- bug reports; bug report classification; N-gram IDF 
\end{abstract}

\section{Introduction}

Bug reports are used for various software development tasks, such as priority and severity assignment \cite{menzies2008automated}, and bug triaging \cite{murphy2004automatic}. 
The quality and reliability of these activities highly depends on the information of bug reports. However, previous studies showed the problem of bug reports can lead to tasks that produce unreliable results. Bettenburg et al. \cite{bettenburg2008makes} investigated the quality of bug reports and found that reports often come with incomplete and incorrect information. As a consequence, developers take much effort on the error inspection process. Antoniol et al. \cite{antoniol2008bug} studied the bug reports misclassification problem; i.e., reports which are labeled as bugs, but actually are non-bug issues. This problem occurs due to the misuse of the bug tracking system (BTS). BTS is used to manage issues related to \textit{bugs}, and is also used for managing issues of other software activities, e.g. request for a new feature, performance improvement, source code refactoring and so on. Due to these various usages of BTS, bug reports are more likely to be misclassified. 

    \par Separating bugs from other request is challenging because each report needs an independent inspection to identify the incorrectness of report. A number of studies \cite{antoniol2008bug,thung2012empirical,herzig2013s} spent much time on manually classify bug reports. Herzig et al. spent 90 days to manually classify over 7,000 bug reports. They found that about one-third of inspected bug reports are actually not the bugs \cite{herzig2013s}. According to their research, manual inspection is difficult and need a lot of effort to be done. For this reason, a technique to automatically classify bug reports is needed.
    
    \par Several studies have attempted to tackle the misclassification of bugs problem. For example, Antoniol et al.  \cite{antoniol2008bug} proposed a word-based automatic classification technique and got decent classification results. Recent studies proposed classification models based on topic modeling techniques; Latent Dirichlet Allocation (LDA)  \cite{pingclasai2013classifying} and Hierarchical Dirichlet Process (HDP) \cite{limsettho2014comparing}. Their proposed techniques are comparable with each other and both of them outperform word-based models in almost all of evaluated cases.
    
    Topic modeling has been regarded as the state-of-the-art Information Retrieval technique and has been applied to various software engineering tasks, such as document clustering, concept/feature location, evolution analysis, uncovering traceability links, defect prediction, searching, and so on. \cite{chen2016survey}. However, it is widely known that although the topic modeling performance strongly depends on model parameters, there is no single recommended method for selecting the number of topics to model \cite{Layman:2016:TMN:2901739.2901760,Panichella:2013:EUT:2486788.2486857}.

    \par In this paper, we propose to apply an alternative technique to classify bug reports, namely, N-gram IDF, a theoretical extension of Inverse Document Frequency (IDF) \cite{shirakawa2015n}. IDF is a numerical statistic of how much information the word provides, that is, IDF is a measure of the rareness of term. It is often used as a weighting factor in information retrieval. However, IDF cannot handle N-grams for N $>$ 1; i.e., phrases that are composed of two or more words. N-gram IDF is capable of handling words and phrases of any length. We can extract all of valid N-gram words and select dominant N-grams by comparing the weight of word and phrases. 

    \par We compare N-gram IDF-based models and topic modeling-based models on three open-source software projects including HTTPClient, Jackrabbit, and Lucene, presented in the previous study \cite{herzig2013s}.

    \par The contribution of this paper can be summarized as follows
\begin{itemize}
\item We propose an automatic classification model based on N-gram IDF-based technique.
\item We evaluate the performance of N-gram IDF-based classification models compared with topic-based models.
\end{itemize}


\section{Methodology}\label{sec:methodology}
%
\begin{figure}
  \centering
  \includegraphics[width=.8\linewidth]{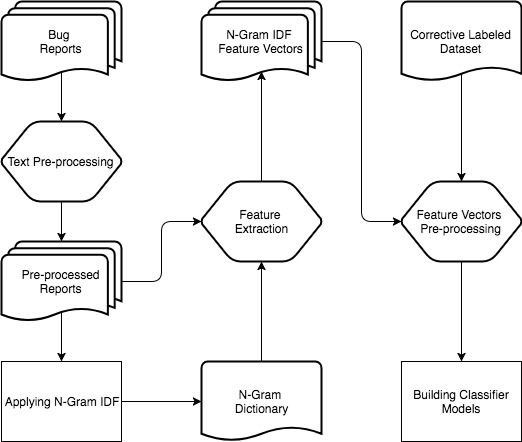}
  \caption{Overview of our automatic bug reports classification model}
  \label{fig:framework}
\end{figure}

\subsection{Overview}
Figure \ref{fig:framework} shows the overall structure of our classification model building process. 
To construct the automatic classification model, firstly, we parse and pre-process the retrieved bug reports files. We then apply the N-gram IDF to the corpora of pre-processed bug report files. The output of this process is a list of all valid N-gram key terms. For each bug report, we then count the raw frequency of each N-gram word and keep the frequency value as a collection of membership vectors. Then, with the retrieved vectors, we combine them with the dataset containing the correct bug report type for each bug report file. Lastly, we use these combined vectors as inputs to train our classification model. 
We explain more detail in the following subsections.

\subsection{Text Processing}



This step has a big impact in optimizing the classification step and data noise removal. With the retrieved bug reports files, we remove some characters related to the programming syntax (e.g., ``=='', ``+'', ``--''). These steps are taken to ensure the quality of our classification models since they carry less meaningful context.

\subsection{Applying N-gram IDF}
By applying N-gram IDF, we can obtain dominant N-gram among overlapping one and extract key terms of any length from a corpus of documents. In this research, we use an N-gram Weighting Scheme tool 
\footnote{https://github.com/iwnsew/ngweight}.
This tool uses enhanced suffix array \cite{Abouelhoda:2004:RST:985384.985389} to enumerate valid N-grams.
The output after applying N-gram IDF tool to the pre-processed data is an N-gram dictionary, which is a list of all valid N-gram key terms. 

\subsection{Feature Extraction}
After we get the N-gram dictionary, we filter out N-gram words that appeared in only one bug report. We then create a feature vector space from the bug report corpus and the N-gram dictionary. For each bug report, we count the raw frequency of each N-gram word and create a vector element based on the raw frequency value. Finally, each vector element contains bug report ID and raw frequency values of all N-gram words that were occurred in the documents. These vector elements will be utilized as features for the next classification phase.


\subsection{Feature Vector Pre-Processing}
Before building classification model, we use the following two feature selection methods to filter out N-gram words that are found to have less impact on classification models.
\begin{enumerate}
\item Correlation-based feature selection: CFS selects a subset of features that are highly correlated with classification while uncorrelated to each other. We use CFS as a feature selection method on 10-fold cross-validation setup by using \textit{Weka}\footnote{http://www.cs.waikato.ac.nz/ml/weka/}.
\item Chi-squared stats: A statistical method assessing the goodness of fit between features and classification. We use Chi-squared stats as a feature selection method on training - testing setup by using Python library \textit{scikit-learn}\footnote{http://scikit-learn.org/stable/index.html}.
\end{enumerate}

\subsection{Building Classifier Model}
As we focus on comparing the performance between N-gram IDF and topic modeling. We separately build up classification models. Each model is constructed from a different combination of text processing techniques and data classification techniques. For the data classification techniques, we use two techniques,  Logistic Regression and Random Forest.

\begin{table}[h!]
\caption{Details of study subjects}
\label{table:study subject}
\centering
 \begin{tabular}{c r r r} 
 \hline
 Project & \# of Reports & \# of Bugs & \# of Other Requests \\
 \hline \\
 HTTPClient & 745 & 305 & 440\\ [1ex]
 Jackrabbit & 2,402 & 938 & 1,464\\ [1ex] 
 Lucene & 2443 & 173 & 1,746\\ [1ex] 
 Cross Project & 5,590 & 1,940 & 3,650\\ [1ex] 
 \hline
\end{tabular}
\end{table}
\begin{table*}[]
\centering
\caption{F-measure comparison table between topic-based and n-gram idf-based classification models by 10-fold cross-validation}
\label{table:fold_result}
\begin{tabular}{|l|c|c|c|c|c|c|c|c|}
\hline
                          & \multicolumn{4}{c|}{\textbf{Logistic Regression}}                                    & \multicolumn{4}{c|}{\textbf{Random Forest}}                                          \\ \cline{2-9} 
                          & \textbf{HTTPClient} & \textbf{Jackrabbit} & \textbf{Lucene} & \textbf{Cross Project} & \textbf{HTTPClient} & \textbf{Jackrabbit} & \textbf{Lucene} & \textbf{Cross Project} \\ \hline
\textbf{Topic-based}      & 0.739               & 0.744               & 0.766           & 0.724                  & 0.721               & 0.717               & 0.756           & 0.712                  \\
\rowcolor[HTML]{9B9B9B} 
\textbf{N-gram IDF-based} & 0.805               & 0.805               & 0.884           & 0.814                  & 0.814               & 0.771               & 0.823           & 0.792                  \\ \hline
\end{tabular}
\end{table*}
\begin{table*}[]
\centering
\caption{F-measure comparison table between topic-based and n-gram idf-based classification models by training - testing setup}
\label{table:train_result}
\begin{tabular}{|l|c|c|c|c|c|c|c|c|}
\hline
                          & \multicolumn{4}{c|}{\textbf{Logistic Regression}}                                    & \multicolumn{4}{c|}{\textbf{Random Forest}}                                          \\ \cline{2-9} 
                          & \textbf{HTTPClient} & \textbf{Jackrabbit} & \textbf{Lucene} & \textbf{Cross Project} & \textbf{HTTPClient} & \textbf{Jackrabbit} & \textbf{Lucene} & \textbf{Cross Project} \\ \hline
\textbf{Topic-based}      & 0.516               & 0.511               & 0.562           & 0.592                  & 0.494               & 0.514               & 0.566           & 0.542                  \\
\rowcolor[HTML]{9B9B9B} 
\textbf{N-gram IDF-based} & 0.687               & 0.646               & 0.731           & 0.658                  & 0.673               & 0.628               & 0.685           & 0.674                  \\ \hline
\end{tabular}
\end{table*}

\section{Experimental Design}\label{sec:Design}
%

\subsection{Study Subjects}
Our research makes use of training data from a previous study \cite{herzig2013s}\footnote{http://www.st.cs.uni-saarland.de/softevo//bugclassify/}. The datasets we gathered are three open-source software projects that use JIRA as an issue tracking system, the name of projects are HTTPClient, Jackrabbit, and Lucene. Each dataset has three fields, i.e. report ID, original type, and its corrected type. We utilize the bug report ID containing in the dataset to retrieve bug report files from software project repository and use corrected type of bug reports as our evaluation baseline. In addition to three open-source software projects datasets, we also created a new dataset by combining all of the bug reports from those three datasets into only one dataset. We call the new dataset as the \textit{Cross Project} dataset. The number of examined bug report files for each project in this study is represented in Table \ref{table:study subject}. We made our final feature vectors dataset available\footnote{https://github.com/sefield/BugReportClassificationWithNgramIDFDataset}.

\subsection{Evaluation Settings}
In this study, we adopt a \textit {F-measure} score as a performance evaluation metric to evaluate our classification models. For our evaluation, we prepare two setups to validate the models. First is 10-fold cross-validation. The idea behind 10-fold cross-validation is that dataset is randomly partitioned into 10 equal size subsets. From these 10 subsets, nine subsets are used as a training data while the other part is retained as a testing data. The process is then repeated 10 times, with all of the subsets are used as a testing data once. We report the average value of F-measure after 10 runs of cross-validation. The other method is a training set and testing set. The idea behind this method is to include a time factor into our classification models to make a scenario more practical, therefore, we split the dataset into a training set and testing set based on reported date of bug reports. The oldest 90\% of bug reports are utilized as a training set while the newest 10\% of bug reports are utilized as a testing set. We also report the value of F-measure from this testing method.
%

We compare the classification performance between N-gram IDF-based models and topic-based models. To build topic-based classification models, we follow the methodology conducted on the previous study \cite{pingclasai2013classifying}. We build a collection of topic membership vectors with 50 as a number of topics, which achieved the highest performance in the previous study. 

\section{Results}\label{sec:result}
In this section, we report the classification performance of proposed N-gram IDF-based models on two testing environments; 10-fold cross-validation and training-testing setups.

In this study, the numbers of N-gram words in the dictionary vary between 58,000 to 530,000. After applying the feature selection methods, the numbers of remaining N-gram words that are used as the features for classification models vary from 40 to 200.

\subsection{10-Fold Cross-Validation Setup}
Table \ref{table:fold_result} shows the evaluation result on 10-fold cross-validation setup. The first column indicates the text processing technique that was used to train the models. For the other columns, four consecutive columns are grouped into a set of study projects. Each set reports the performance of a classification model that employed one of the two classifier algorithms. As we see in Table \ref{table:fold_result}, N-gram IDF-based models outperform topic-based models in all of cases with the F-measure score varies between  0.80 - 0.81, 0.77 - 0.80, 0.82 - 0.88, and 0.79 - 0.81 for HTTPClient, Jackrabbit, Lucene, and Cross Project respectively.

\subsection{Training - Testing Setup}
Table \ref{table:train_result} shows the evaluation result on training - testing setup. The first column indicates the text processing technique that was used to train the models. For the other columns, four consecutive columns are grouped into a set of study projects. Each set reports the performance of a classification model that employed one of the two classifier algorithms. As we see in Table \ref{table:train_result}, N-gram IDF-based models perform better than the topic-based models with the F-measure score varies between 0.67 - 0.69, 0.62 - 0.65, 0.68 - 0.73, and 0.65 - 0.67 for HTTPClient, Jackrabbit, Lucene, and Cross Project respectively.
	
\begin{figure}[ht]
\centering     
\subfigure[HTTPClient]{\label{fig:httpclient_boxplot}\includegraphics[width=0.49\linewidth]{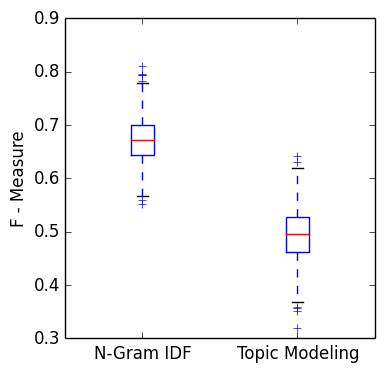}}
\subfigure[Jackrabbit]{\label{fig:jackrabbit_boxplot}\includegraphics[width=0.49\linewidth]{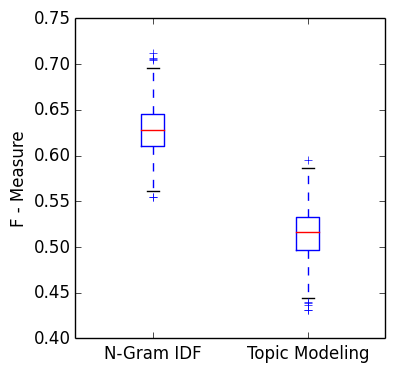}}
\subfigure[Lucene]{\label{fig:lucene_boxplot}\includegraphics[width=0.49\linewidth]{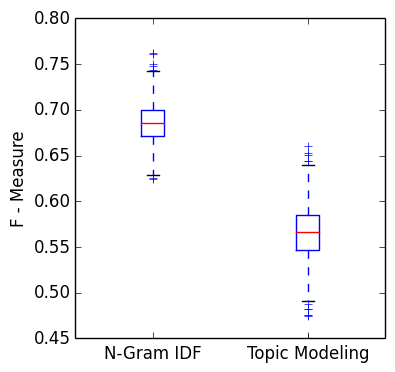}}
\subfigure[Cross Project]{\label{fig:cross_project_boxplot}\includegraphics[width=0.49\linewidth]{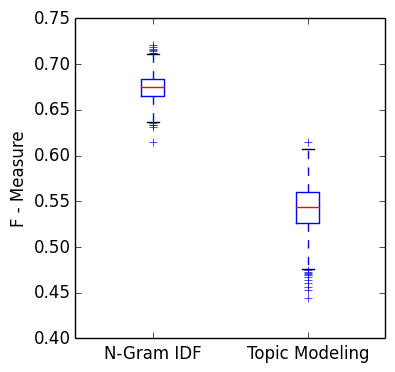}}
\caption{Boxplots of N-gram IDF-based models and Topic-based models. F-measure value on a 1,000 times run of Random Forest}
\label{fig:boxplot}
\end{figure}

According to Arcuri and Briand \cite{arcuri2011practical},  it is recommended to use a high number of runs to assess the results of randomized algorithms because we get a different result on every run while applying the algorithms to the same problem instance. Following the study guideline \cite{arcuri2011practical}, we conducted 1,000 runs of random forest for both N-gram IDF-based and topic-based models.

	\par Figure \ref{fig:boxplot} shows the results of the 1,000 runs of random forest on the training - testing setup. For each study project, boxplots of the value of F-measure for N-gram IDF-based model and topic-based model are shown. In all study projects, N-gram IDF-based models achieved higher performance than topic-based models. The Mann-Whitney U-test was applied to detecting statistical differences between N-gram IDF-based model and topic-based model. We found that in all of the study projects, the differences are statistically significant (p-value $<$ 0.001).

\section{Discussion}\label{sec:discussion}

\subsection{Why N-gram IDF Works Well?}

    Table \ref{table:n_gram_words} shows examples of N-gram words that were identified and used as features for classification process, along with their implication, global term frequency (gtf, an amount of occurred N-gram word in all of document.) and document frequency (df, amount of documents that have the N-gram word.). Since N-gram IDF can extract all of valid N-gram words of any length, it is capable of capturing the text structure in a sequential order. As we see in Table \ref{table:n_gram_words}, the N-gram IDF can extract the key terms that vary in both contexts and lengths. For example, \textit{nullpointerexception} and \textit{how to reproduce} intuitively indicate BUG, while \textit{improvement}  infers that the report is not related to BUG. Therefore, these key terms extracted from N-gram IDF can be used as the features that contribute to the efficiency of our classification models.
    With the strong point of N-gram IDF, we believe that our proposed N-gram IDF classification technique can be further developed not only to be used on the bug report classification task, but also on the other software engineering tasks, such as duplicate bug report detection and bug localization by linking code and bug reports.
   \par Technically, a time that our technique take to create and train a classification model depends on the size of document corpus. However, in a case of all required tools and dataset are provided with all of the necessary variables, we can complete the whole process starting from the scratch within a few hours. Therefore, our technique can help developers to accelerate their development process by reducing their time on bug reports inspection process as they do not need to waste much effort and time on the manual bug report classification process.

\begin{table}
\centering
\caption{Example of extracted N-grams on Cross Project Dataset (gtf = global term frequency, df = document frequency)}
\label{table:n_gram_words}
\begin{tabular}{|l|l|r|r|}
\hline
\rowcolor[HTML]{C0C0C0} 
N-gram                                                             & Implication                                                                                                 & gtf & df  \\ \hline
nullpointerexception                                               & \begin{tabular}[c]{@{}l@{}}Reports with a null pointer \\exception error is related to BUG.\end{tabular} & 252 & 140 \\ \hline
how to reproduce                                                   & \begin{tabular}[c]{@{}l@{}}Reports with reproduce step \\ should related to BUG.\end{tabular}               & 17  & 14  \\ \hline
improvement                                                        & \begin{tabular}[c]{@{}l@{}}Improvement discussion \\ should not related to BUG.\end{tabular}   & 464 & 295 \\ \hline
\begin{tabular}[c]{@{}l@{}}demonstrate the \\ problem\end{tabular} & \begin{tabular}[c]{@{}l@{}}Discussion with demo of problem\\ should related to BUG\end{tabular}             & 11  & 9   \\ \hline
performance test                                                   & \begin{tabular}[c]{@{}l@{}}Performance test discussion \\ should not related to BUG\end{tabular}             & 50  & 32  \\ \hline
\end{tabular}
\end{table}
\subsection{Threats to Validity}
    \textbf{Labeled dataset depends on the previous study.}
This is a threat to construct validity.
Although the data are inspected with fixed rules, errors might still occur. Moreover, the rules for manual classification also depend on an individual perspective. If the different rules were employed, our classification models would produce different results.
    \par \textbf{Study subjects are only open source projects.} 
This limitation is a threat to external validity.
All bug reports in our datasets are written in Java and use the JIRA bug tracker, which might not be generalized to the projects that are written in other programming languages or using other bug tracking systems.
\section{Related Work}\label{sec:related}

    Bug reports are essential software artifacts for software projects, especially in open-source software projects. According to bug-report analysis survey paper \cite{zhang2015survey}, many amounts of research have been conducted on bug-report analysis from many aspects due to an availability of a lot of bug reports, such as predicting the severity of bug reports \cite{menzies2008automated}, and bug reports triaging \cite{murphy2004automatic}. With these kinds of tasks, quality of bug reports is important and begin normally concerned. Among a number of studies so far, there are some of them that focus on the issue of misclassification of bug reports \cite{antoniol2008bug,herzig2013s,pingclasai2013classifying,limsettho2014comparing}.

    \par Antoniol et al. \cite{antoniol2008bug} first introduced the bug reports misclassification problem. They found that less than half of bug reports are actually related to bugs. This shows the complex usage of Bug Tracking System, which causes bug reports misclassification. They also proposed a text-based automated classification method. They concluded that the information contained in bug reports can be indeed used to classify bugs from other activities. Herzig et al. \cite{herzig2013s} manually inspected 7,401 issue reports to learn the percentage of bug reports that have been misclassified and found that ``Every third bug is not a bug.'' They indicated the possible impacts and bias in bug report prediction models occurred from the misclassification problem. However, manual inspection process needs high effort. Pingclasai et al. \cite{pingclasai2013classifying} addressed the issue and proposed another approach to automatically classifying bug reports by using one of topic modeling techniques; Latent Dirichlet Allocation (LDA). According to their results, naive Bayes classifier is the most efficient classification model when applying LDA. Limsettho et al. \cite{limsettho2014comparing} also addressed the issue and proposed another automated classification method by applying another topic modeling techniques; Hierarchical Dirichlet Process (HDP). They found that HDP performance is comparable with the previous study, but without parameter tuning. Zhou et al. \cite{zhou2016combining} also address the issue and proposed a hybrid approach by combining both text mining and data mining techniques via a technique called data grafting. 
    In this work, we address the bug reports misclassification issue and propose alternative automated classification approach by applying N-gram IDF.

\section{Conclusions}\label{sec:conclusions}
In this paper, we propose a method for automatically classifying bug reports based on textual information contained in each report. N-gram IDF was adopted to extract key terms from corpora of bug reports. Our technique aims to reduce time and effort required for manual inspection. Our experiment conducted on three open-source software projects (HTTPClient, Jackrabbit, and Lucene). We built classification models separately from two classification algorithms, i.e. logistic regression and random forest. Based on the results, we conclude that
\begin{itemize}
\item Collection of key N-gram words extracted from a corpus of bug reports by applying N-gram IDF is able to separate bugs from another request on other software activities. Our classification models have an F-measure score between 0.62 and 0.88. 
\item Both logistic regression and random forest model have similar performance range. Both seem to have an opportunity to be further developed to achieve higher classification performance.
\end{itemize}

    \par Moreover, we also compare classification performance between N-gram IDF-based model and topic-based model. The result shows that N-gram IDF-based model outperforms topic-based model in all of the evaluated cases. 
    
    \par For future work, We plan to improve the performance of current classification method. We also plan to extend our work into the classification model of multiclass label corpus. We also want to generalize our result by experimenting on different BTS and other software projects written in other programming languages. Lastly, we aim to conduct an experiment not only on bug reports classification but also on other software engineering activities.

\section*{Acknowledgment}
This work has been supported by JSPS KAKENHI Grant Number 16H05857.


\begin{thebibliography}{10}

\bibitem{menzies2008automated}
T.~Menzies and A.~Marcus, ``Automated severity assessment of software defect
  reports,'' in {\em Proceedings of IEEE International Conference on Software
  Maintenance (ICSM)}, pp.~346--355, IEEE, 2008.

\bibitem{murphy2004automatic}
G.~Murphy and D.~Cubranic, ``Automatic bug triage using text categorization,''
  in {\em Proceedings of 16th International Conference on Software Engineering
  \& Knowledge Engineering (SEKE)}, 2004.

\bibitem{bettenburg2008makes}
N.~Bettenburg, S.~Just, A.~Schr{\"o}ter, C.~Weiss, R.~Premraj, and
  T.~Zimmermann, ``What makes a good bug report?,'' in {\em Proceedings of 16th
  ACM SIGSOFT International Symposium on Foundations of software engineering
  (FSE)}, pp.~308--318, ACM, 2008.

\bibitem{antoniol2008bug}
G.~Antoniol, K.~Ayari, M.~Di~Penta, F.~Khomh, and Y.-G. Gu{\'e}h{\'e}neuc, ``Is
  it a bug or an enhancement?: a text-based approach to classify change
  requests,'' in {\em Proceedings of 2008 conference of the center for advanced
  studies on collaborative research: meeting of minds (CASCON)}, p.~23, ACM,
  2008.

\bibitem{thung2012empirical}
F.~Thung, S.~Wang, D.~Lo, and L.~Jiang, ``An empirical study of bugs in machine
  learning systems,'' in {\em Proceedings of IEEE 23rd International Symposium
  on Software Reliability Engineering (ISSRE)}, pp.~271--280, IEEE, 2012.

\bibitem{herzig2013s}
K.~Herzig, S.~Just, and A.~Zeller, ``It's not a bug, it's a feature: how
  misclassification impacts bug prediction,'' in {\em Proceedings of 2013
  International Conference on Software Engineering (ICSE)}, pp.~392--401, IEEE
  Press, 2013.

\bibitem{pingclasai2013classifying}
N.~Pingclasai, H.~Hata, and K.~Matsumoto, ``Classifying bug reports to bugs and
  other requests using topic modeling,'' in {\em Proceedings of 20th
  Asia-Pacific Software Engineering Conference (APSEC)}, vol.~2, pp.~13--18,
  IEEE, 2013.

\bibitem{limsettho2014comparing}
N.~Limsettho, H.~Hata, and K.~Matsumoto, ``Comparing hierarchical dirichlet
  process with latent dirichlet allocation in bug report multiclass
  classification,'' in {\em Proceedings of 15th IEEE/ACIS International
  Conference on Software Engineering, Artificial Intelligence, Networking and
  Parallel/Distributed Computing (SNPD)}, pp.~1--6, IEEE, 2014.

\bibitem{chen2016survey}
T.-H. Chen, S.~W. Thomas, and A.~E. Hassan, ``A survey on the use of topic
  models when mining software repositories,'' {\em Empirical Software
  Engineering}, vol.~21, no.~5, pp.~1843--1919, 2016.

\bibitem{Layman:2016:TMN:2901739.2901760}
L.~Layman, A.~P. Nikora, J.~Meek, and T.~Menzies, ``Topic modeling of nasa
  space system problem reports: Research in practice,'' in {\em Proceedings of
  13th International Conference on Mining Software Repositories (MSR)},
  pp.~303--314, ACM, 2016.

\bibitem{Panichella:2013:EUT:2486788.2486857}
A.~Panichella, B.~Dit, R.~Oliveto, M.~Di~Penta, D.~Poshyvanyk, and A.~De~Lucia,
  ``How to effectively use topic models for software engineering tasks? an
  approach based on genetic algorithms,'' in {\em Proceedings of 2013
  International Conference on Software Engineering (ICSE)}, pp.~522--531, IEEE
  Press, 2013.

\bibitem{shirakawa2015n}
M.~Shirakawa, T.~Hara, and S.~Nishio, ``N-gram idf: A global term weighting
  scheme based on information distance,'' in {\em Proceedings of 24th
  International Conference on World Wide Web (WWW)}, pp.~960--970, ACM, 2015.

\bibitem{Abouelhoda:2004:RST:985384.985389}
M.~I. Abouelhoda, S.~Kurtz, and E.~Ohlebusch, ``Replacing suffix trees with
  enhanced suffix arrays,'' {\em J. of Discrete Algorithms}, vol.~2,
  pp.~53--86, Mar. 2004.

\bibitem{arcuri2011practical}
A.~Arcuri and L.~Briand, ``A practical guide for using statistical tests to
  assess randomized algorithms in software engineering,'' in {\em Proceedings
  of 33rd International Conference on Software Engineering (ICSE)}, pp.~1--10,
  IEEE, 2011.

\bibitem{zhang2015survey}
J.~Zhang, X.~Wang, D.~Hao, B.~Xie, L.~Zhang, and H.~Mei, ``A survey on
  bug-report analysis,'' {\em Science China Information Sciences}, vol.~58,
  no.~2, pp.~1--24, 2015.

\bibitem{zhou2016combining}
Y.~Zhou, Y.~Tong, R.~Gu, and H.~Gall, ``Combining text mining and data mining
  for bug report classification,'' {\em Journal of Software: Evolution and
  Process}, vol.~28, no.~3, pp.~150--176, 2016.

\end{thebibliography}

\end{document}